\documentclass[aps,prl,reprint,groupedaddress]{revtex4-2}
\usepackage{graphicx}
\usepackage{amsmath, amsthm, amssymb, amsfonts, bm}

\usepackage{color} 
\usepackage[normalem]{ulem}
\usepackage{hyperref}

\hypersetup{
    colorlinks=true,
    linkcolor=blue,
    filecolor=magenta,      
    urlcolor=blue,
}

\begin{document}

%Title of paper
\title{On the accessibility of stable reactor operating regimes in quasi-symmetric stellarators}

\author{A. M. Wright}
\email[]{adelle.wright@wisc.edu}
\author{B. J. Faber}
\affiliation{University of Wisconsin - Madison}

\date{\today}

\begin{abstract}
Maximizing particle and energy confinement is crucial for achieving the sustained burning plasma conditions necessary to realize fusion energy.
For stellarator reactors, one proposed strategy for avoiding destructive instabilities is to operate at high-field but low(er) plasma pressure.
In this work, we investigate the accessibility of such a reactor-relevant low-beta regime in a reactor-scale quasi-axisymmetric stellarator using state-of-the-art high-fidelity macro- and microscopic simulation tools.
We consider a configuration with a flattened core pressure profile and favourable properties from the macroscopic and neoclassical perspectives.
By contrast, linear and nonlinear calculations with the \textsc{Gene} code show an abrupt transition to a regime of highly deleterious transport at low (local) plasma beta.
We describe the characterisation of these transport regimes as well as the confinement transition.
We discuss the implications broadly for stellarator optimisation and highlight the impact on quasi-symmetric stellarator design strategies.
\end{abstract}

\maketitle
%%%%%%%%%%%%%%%%%%%%%%%%%%%%%%%%%%%%%%%%%%%%%%%%%%%%%%%%%%%%%%%%%%%%%%%%%%%%%%%%%%%%%%%%%%%%%%%%%%%%%%%%%%%%%%%%%%%%%%%%%%%%%%%%%%%%%%%%%%%%
\section{Introduction}\label{sec:1}
Stellarators have great potential for realising fusion energy via toroidal confinement of high-temperature, magnetised plasmas.
Since these configurations are intrinsically non-axisymmetric (i.e.\ three-dimensional), there is considerable flexibility for optimising the plasma shape, to achieve desirable physics or engineering properties \citep{imbert-gerard2024}.
Indeed, numerous candidate configurations have been proposed in both the public \citep{beidler2001,najmabadi2008} and private sectors \citep{swanson2025,hegna2025,lion2025}, using recently developed computational code suites \citep{landreman2021,panici2023}.
Confinement of stellarator plasmas is achieved almost entirely through geometric shaping with external, electromagnetic coils.
As such, stellarators are expected to have greater intrinsic robustness to macroscopic, magnetohydrodynamic (MHD) instabilities.
This expectation is supported by experiments at the Large Helical Device (LHD), which have shown that it is possible to exceed linear stability thresholds, without detrimentally impacting energy confinement \citep{sakakibara2006,watanabe2005}.
Moreover, simulations of LHD plasmas have shown that high wave number MHD instabilities require coupling to low-$n$ instabilities, where $n$ is the toroidal mode number, in order to trigger large-scale loss-of-confinement \citep{civit-bertran2025,mizuguchi2009}.
In this absence of this type of mode coupling, it begs the question; how important is linear MHD stability for high-performance stellarator plasmas?
If the answer is negative, then this is highly advantageous for stellarator optimisation since it has been shown that enforcing linear MHD stability is antagonistic to desirable coil characteristics \citep{feng2020}.

At the microscopic level, it has now been conclusively demonstrated that neoclassical transport losses can be very effectively suppressed with modern optimisation strategies \citep{landreman2022}.
As such, drift wave turbulence is expected to be the dominant source of heat and particle losses in future stellarator reactors \citep{hegna2022}.
Specifically, the ion temperature gradient (ITG) instability and kinetic ballooning mode (KBM).
ITG is stabilised with increasing plasma pressure ($\beta$) \citep{pueschel2010, aleynikova2018} and so, while it may modify the pathway to a finite-$\beta$ operating point, significant progress has been made in understanding so-called `ion temperature clamping' \citep{beurskens2021} and ITG suppression through optimisation \citep{hegna2018, kim2024, landreman2025}.
By contrast, KBMs are destablised with increasing plasma $\beta$ \citep{aleynikova2018, mckinney2021} and computational studies of Wendelstein 7-X have suggested KBMs as the determinant of $\beta$-limits in quasi-isodynamic stellarators \citep{mulholland2023, mulholland2025}.

Future stellarator reactors are proposed to operate at much higher magnetic field strength than current devices \citep{swanson2025, hegna2025, lion2025}.
By significantly reducing the particle gyroradius, this is one promising strategy for reducing turbulent transport.
Irrespective of the target operating point -- whether low-$\beta$ as in some quasi-isodynamic scenarios \citep{guttenfelder2025} or high-$\beta$ as in quasi-axisymmetry \citep{swanson2025} -- understanding the accessibility of high-field, low-$\beta$ regimes is crucial for realising fusion with stellarators. 
In this work, we use high-fidelity simulations to characterise the transport and confinement properties of a high-field, low-$\beta$ reactor scenario from both macro- and microscopic perspectives.
This reveals an abrupt transition in global confinement properties and demonstrates the importance subdominant modes and nonlinear mode excitation.
With this in mind, we highlight the critical need to develop effective techniques for optimising with respect to KBMs when targeting finite-$\beta$ design points.

%%%%%%%%%%%%%%%%%%%%%%%%%%%%%%%%%%%%%%%%%%%%%%%%%%%%%%%%%%%%%%%%%%%%%%%%%%%%%%%%%%%%%%%%%%%%%%%%%%%%%%%%%%%%%%%%%%%%%%%%%%%%%%%%%%%%%%%%%%%%
\section{Method}\label{sec:2}
\begin{figure}
    \centering
    \includegraphics[width=1.0\linewidth]{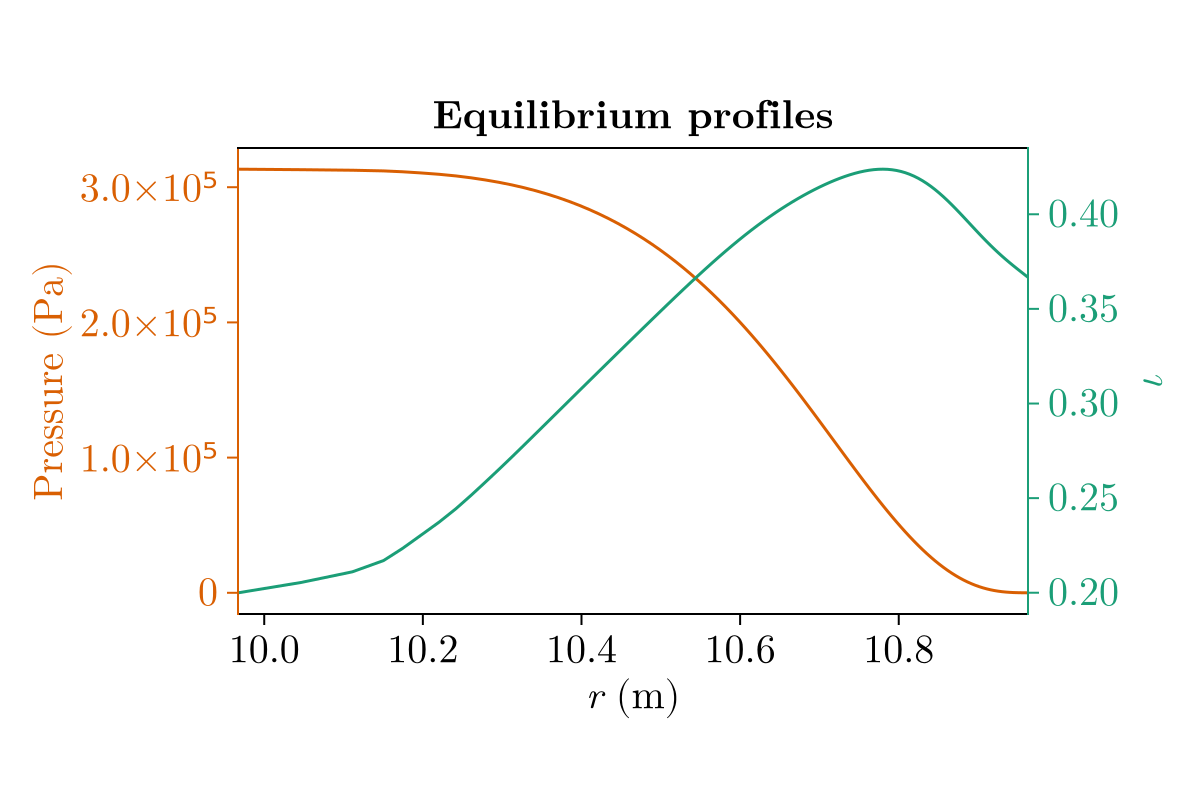}
    \caption{Pressure (left, orange) and rotational transform (right, green) profiles for the quasi-axisymmetric equilibrium under consideration.}
    \label{fig:1}
\end{figure}

In this work, we consider a $\beta=1\%$ three field-period, quasi-axisymmetric equilibrium that was optimised for linear ideal MHD stability, low neoclassical transport (via $\epsilon_{eff}^{1/2}$) and self-consistent bootstrap current as computed by the SFINCS code \citep{feng2020}.
This configuration is one of a series ranging from $\beta=1\% - 5\%$ that was developed as part of a systematic study of plasma elongation under reactor-relevant conditions \citep{feng2020}.
The equilibrium pressure and rotational transform profiles are shown in Figure \ref{fig:1}.
Of particular note is the pressure profile which is relatively flat core and has a steep gradient towards the plasma edge ($s\sim0.7$).
Such a macroscopic profile would be consistent with transport barrier formation \citep{fujisawa2002}.
This aspect ratio 6 configuration has average minor radius $1.55~\text{m}$, on-axis magnetic field $5.7~\text{T}$ and a plasma volume of $444~\text{m}^{3}$. 

To characterise the macro- and microscopic transport properties, we use the M3D-C1 \citep{jardin2012} and \textsc{Gene} \citep{jenko2000} codes, respectively.
M3D-C1 is an extended-MHD code that has been widely used to simulate the macroscopic evolution of toroidally confined fusion plasmas in both tokamak and stellarator geometry.
In this work, we use a visco-resistive single-fluid MHD model.
This includes a temperature-dependent Spitzer resistivity model (rather than a uniform resistivity coefficient as implemented elsewhere \citep{todo2010}), which is important since there is evidence to suggest that this can non-trivially modify the nonlinear evolution of MHD instabilities \citep{wright2024}.
This work predates the bootstrap current modelling capability of M3D-C1, which has since become available \citep{saxena2025}, even so, the microscopic analysis indicates that the timescale for evolution due to bootstrap current is long compared to the timescales of interest in this work.
For the M3D-C1 simulations, we use $2.9\times10^5$ 3D elements and over $\sim 5\times10^5$ CPU-hours, with Spitzer resistivity ($\sim T_3^{-3/2}$) and a normalised viscosity of $\nu=10^{-4}~\text{kg/ms}$, which is consistent with parameters used in previous nonlinear stellarator MHD simulations \citep{wright2024}.

The \textsc{Gene} code solves a gyrokinetic model to simulate micro-instability and micro-turbulence in magnetically confined plasmas for fusion \citep{jenko2000}.
Each plasma species, $s$, is evolved in time and coupled self-consistently with external and internally generated electromagnetic fields via the coupled Vlasov-Maxwell system.
\textsc{Gene} is widely used in both tokamaks and stellarators and has extensive capabilities. 
It models the nonlinear kinetic dynamics for an arbitrary number of ion species (including impurities) and electrons, electrostatic and electromagnetic fluctuation fields, interfaces with MHD codes for providing realistic magnetic field geometry \citep{xanthopoulos2009}, as well as inter- and intra-species collision operators \citep{crandall2020}.
In this work, we use the local version of \textsc{Gene} \citep{jenko2000}, which uses a standard flux-tube computational domain \citep{beer1995} with generalized twist-and-shift boundary conditions \citep{martin2018}.
An interface to the GVEC code \citep{banonnavarro2020} is used to compute the necessary geometric quantities at the $\psi_t/\psi_{edge} = 0.5$ flux surface from the global MHD equilibrium, where $\psi_t$ is the toroidal magnetic flux.

Both linear and nonlinear \textsc{Gene} simulations were performed, the former to compute linear eigemodes and eigenvectors, the latter to assess the nonlinear turbulent transport.
For the linear simulations, a $n_{k_x} \times n_z \times n_{v_\parallel} \times n_{\mu} \times n_s = 3 \times 192 \times 32 \times 8 \times 2$ grid was used, where $\left(n_x,n_z,n_{v_\parallel},n_{\mu}\right)$ are the resolutions in the radial mode number, parallel, parallel velocity, and magnetic moment coordinates, respectively, and $n_s$ is the number of gyrokinetic species.
Nonlinear simulations, a grid with size $n_{k_x} \times n_{k_y} \times n_z \times n_{v_\parallel} \times n_{\mu} \times n_s = 128 \times 64 \times 192 \times 32 \times 8 \times 2$ where used.
Here $n_{k_y}$ is the number binormal wavenumbers used in the simulation, where $\nabla y = \nabla z \times \nabla x$ for the basis vectors $\nabla x$ and $\nabla z$ in the radial and parallel direction, respectively.
In the flux-tube domain, the values of the background plasma temperature and density, as well as their gradients, are free parameters.
To simulate a potential reactor-like scenario, the reference value of the temperature is chosen as $T_\text{ref}=3$ keV, the normalized electron and ion temperature gradients are set to $a/L_{T\mathrm{e}} = a/L_{T\mathrm{i}} = 3$, and the normalized density gradient is set to $a/L_n = 1$.
Here, $a/L_{a\alpha} = -\frac{a}{\alpha}\frac{d\alpha}{dx}$ is the normalized scale length for quantity $\alpha$, $x$ is the \textsc{Gene} radial coordinate, and $a$ is the average minor radius.
The value of the reference density $n_{\text{ref}} = n_0 10^{20}/m^3$ is varied, such that $\beta_\text{e} = 8\pi n_\text{ref} T_{ref}/B_{ref}^2$ is also varied consistently between $0$ and $2\%$.
All simulations are run with $A_{\parallel}$ fluctuations enabled, unity species temperature ratio $T_\mathrm{e}/T_\mathrm{i} = 1$, and no collisions.

%%%%%%%%%%%%%%%%%%%%%%%%%%%%%%%%%%%%%%%%%%%%%%%%%%%%%%%%%%%%%%%%%%%%%%%%%%%%%%%%%%%%%%%%%%%%%%%%%%%%%%%%%%%%%%%%%%%%%%%%%%%%%%%%%%%%%%%%%%%%
\section{Results and discussion}\label{sec:3}
\begin{figure}
    \centering
    \includegraphics[width=1.0\linewidth]{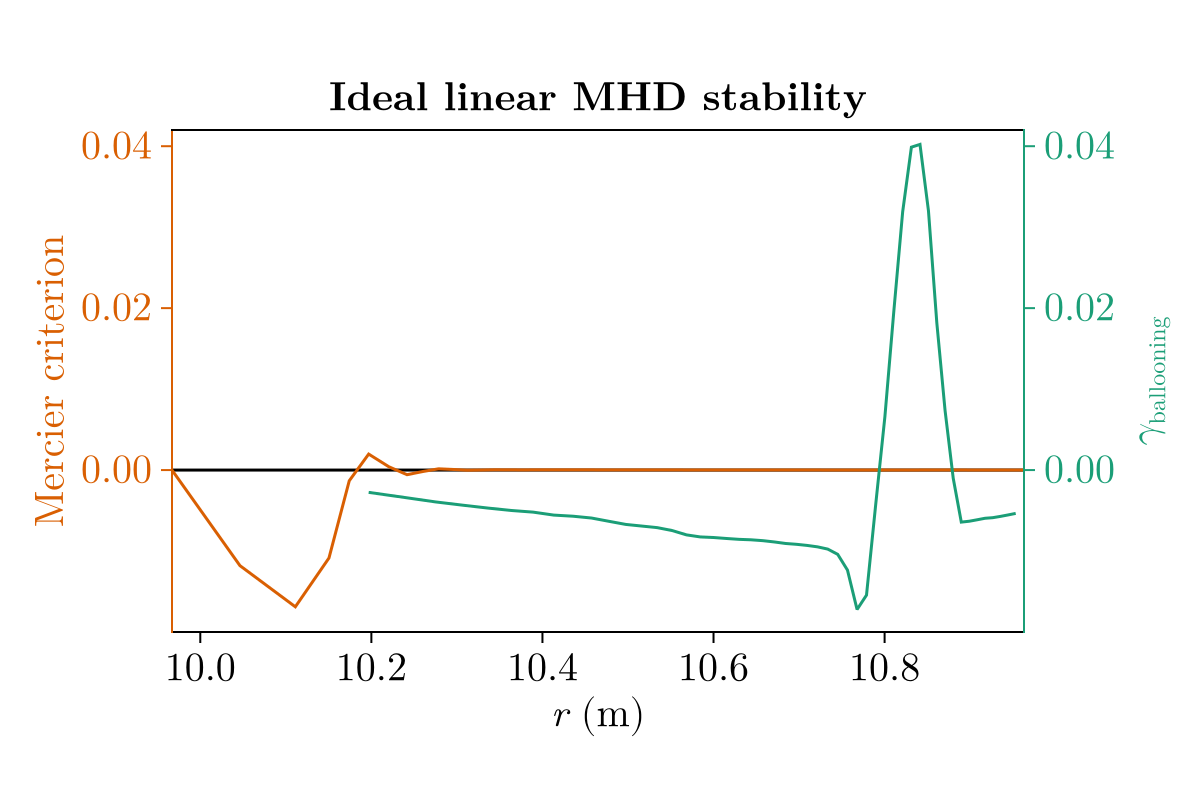}
    \caption{Mercier criterion as computed by VMEC (left, orange) and ballooning growth rate (right, green) for the quasi-axisymmetric equilibrium under consideration.}
    \label{fig:2}
\end{figure}

From the macroscopic perspective, the equilibrium of interest is stable with respect to external modes \citep{feng2020} and satisfies the Mercier criterion, except for a small region in the core.
It is, however, unstable to ballooning modes near the plasma edge, as see Figure \ref{fig:2}.
With a series of nonlinear M3D-C1 simulations, we verify that the equilibrium is indeed unstable to a high-$n$ ballooning mode, which the develops on an ideal timescale ($\sim 100~\tau_A$ where $\tau_A$ is the Alfv\'en time).
Nonlinearly, however, this instability is quite benign and saturates with low amplitude.
It leads to the formation of a narrow magnetic island chain near the plasma edge but otherwise does not drive a significant change in the overall confinement characteristics.
This can be understood from the absence of nonlinear coupling between multiple instabilities which, in other work, was shown to be detrimental \citep{wright2024, civit-bertran2025}.
Thus, we conclude that the macroscopic properties of this equilibrium are desirable from a fusion reactor scenario design standpoint.

An assessment of the linear microinstabilities and nonlinear transport, however, shows quite the opposite.
From the perspective micro-turbulence, the equilibrium has extremely poor confinement properties.
We observe a sudden transition from moderate ITG-dominated transport to a regime of explosive KBM growth, in both the linear and nonlinear simulations.
We find that this transition occurs at very low local plasma-$\beta$ (between $0.67\%$ and $0.85\%$) indicating that the macroscopic state would not be reachable in practice, contradicting the picture provided by the macroscopic analysis.

\begin{figure}
    \centering
    \includegraphics[scale = 0.55]{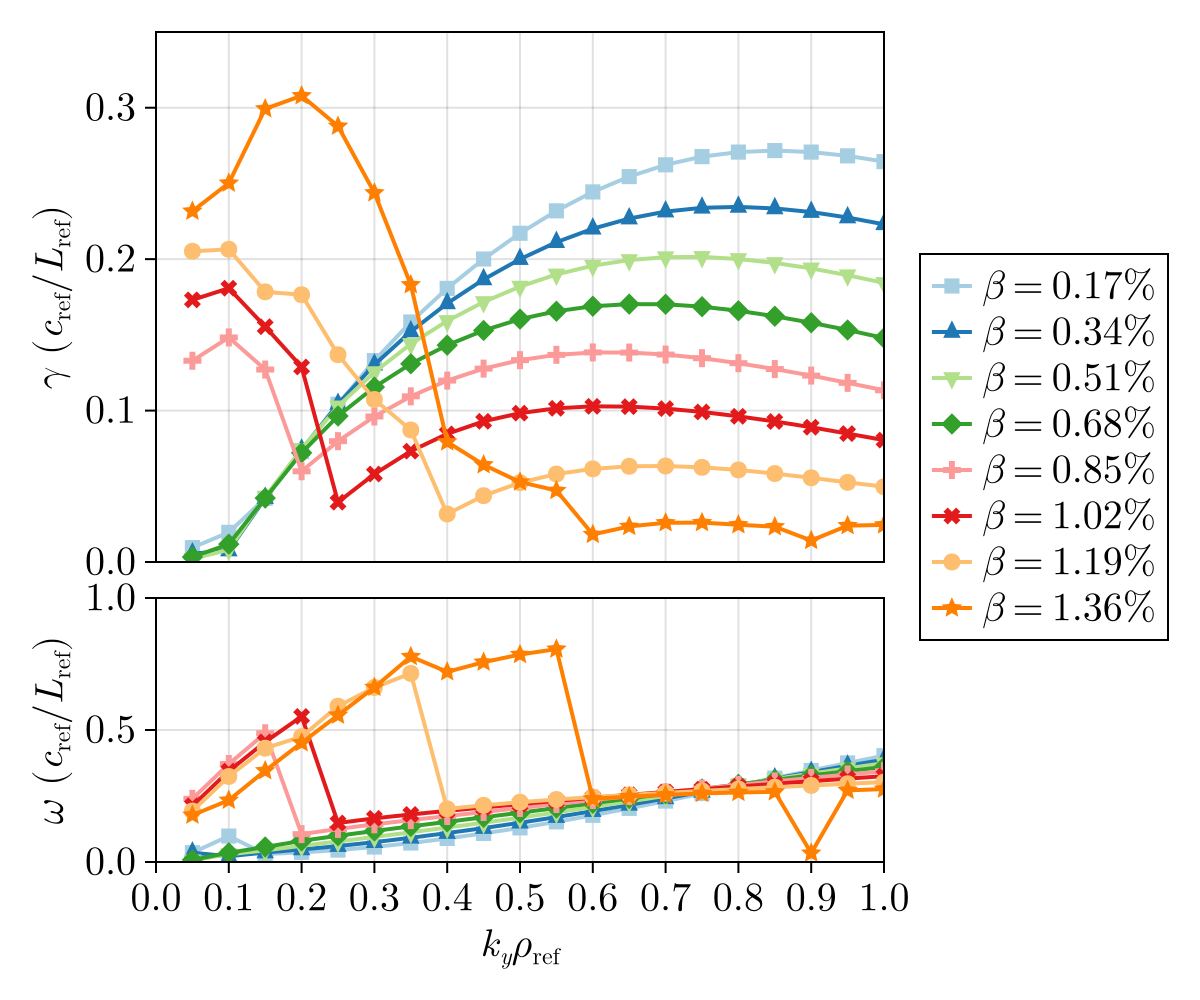}
    \caption{Most unstable linear mode growth rates and frequencies from linear \textsc{Gene} simulations as a function of $\beta$.}
    \label{fig:3}
\end{figure}

In the linear analysis, we see a clear and abrupt bifurcation in the microinstabilty behavior as a function of local plasma-$\beta$.
This is shown in Figure \ref{fig:3}, where the most unstable linear mode growth rates and frequencies are presented as a function of binormal wavenumber $k_y$ for eight different $\beta$ values ranging from $0.17\%$ to $1.36\%$.
At extremely low $\beta$ ($\leq 0.68\%$), the linear growth rates and frequencies are indicative of an ITG being the most unstable mode for all $k_y \leq 1$.
This agrees with previous linear studies of low-$\beta$ stellarators \citep{mckinney2021}.
Moreover, the linear growth rates are suppressed as the $\beta$ fraction increases, which is similar to what is observed for simulations of ITGs in axisymmetric geometry \citep{pueschel2010, aleynikova2018}.

\begin{figure}
    \centering
    \includegraphics[width=1.0\linewidth]{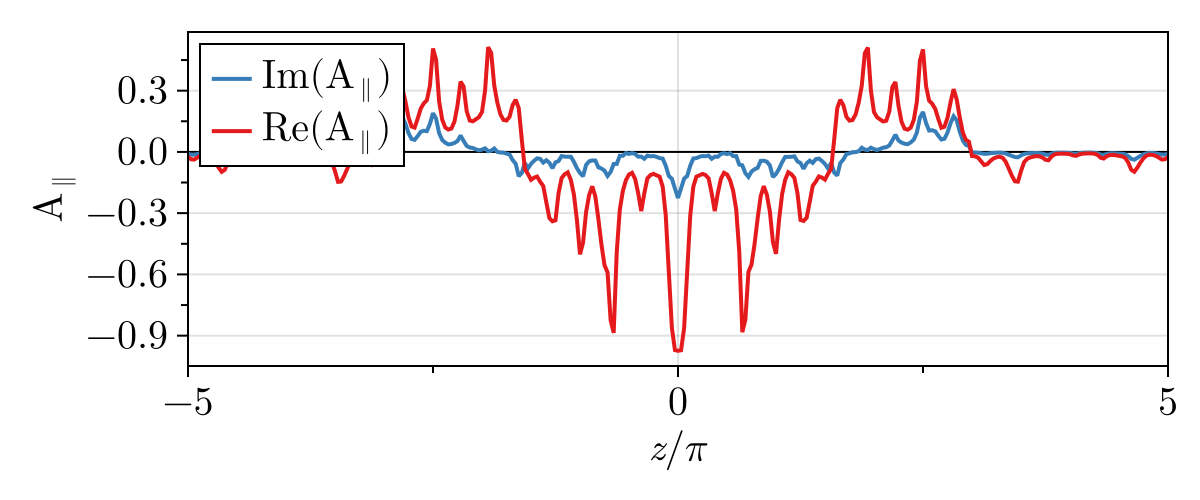}\\
    \includegraphics[width=1.0\linewidth]{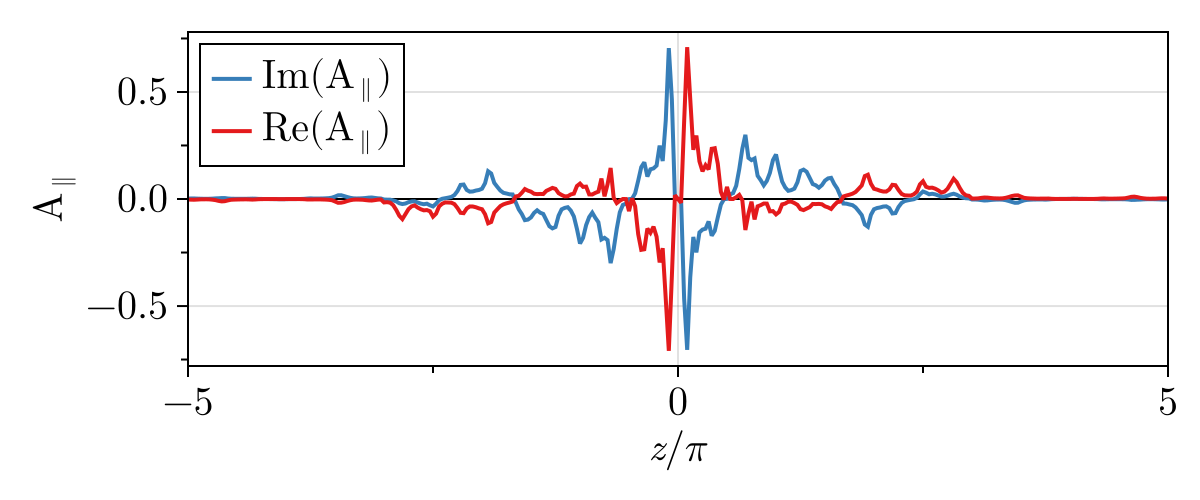}
    \caption{Eigenfunctions of $A_\parallel$ at $k_y = 0.05$  from linear \textsc{Gene} calculations for $\beta = 0.85\%$ (top) and $\beta=0.68\%$ (bottom).}
    \label{fig:4}
\end{figure}

However, at a critical point between $\beta = 0.68\%$ and $\beta = 0.85\%$, there is dramatic jump in both growth rates and frequencies at low $k_y$, indicating a transition from ITG turbulence to a different dominant instability.
This new mode has the signature of a KBM instability as it scales proportionally with the $\beta$ fraction and has even parity in $A_\parallel$, which can be seen from the eigenmode shape in  Figure \ref{fig:4}.
Critically, destabilization of the KBM occurs at a $\beta$ value significantly below the $3\%$ $\beta$-limit for destructive macroscopic MHD instabilities which the configuration was optimized to avoid \citep{feng2020}.
This contrasts with previous studies of KBM limits in quasi-axisymmetric stellarators which found that the KBM destabilization threshold matched the MHD stability $\beta$ limit for NCSX \citep{mckinney2021}.
It does, however, agree with the trends reported in \citep{mulholland2023, mulholland2025}, where KBMs were destabilized in the quasi-isodynamic W7-X stellarator well below the operational (macroscopic) $\beta$ limit.
This suggests the existence of qualitatively different KBM characteristics even in the same class of stellarator configuration, the underlying reason for which is not understood.

\begin{figure}
    \centering
    \includegraphics[width=1.0\linewidth]{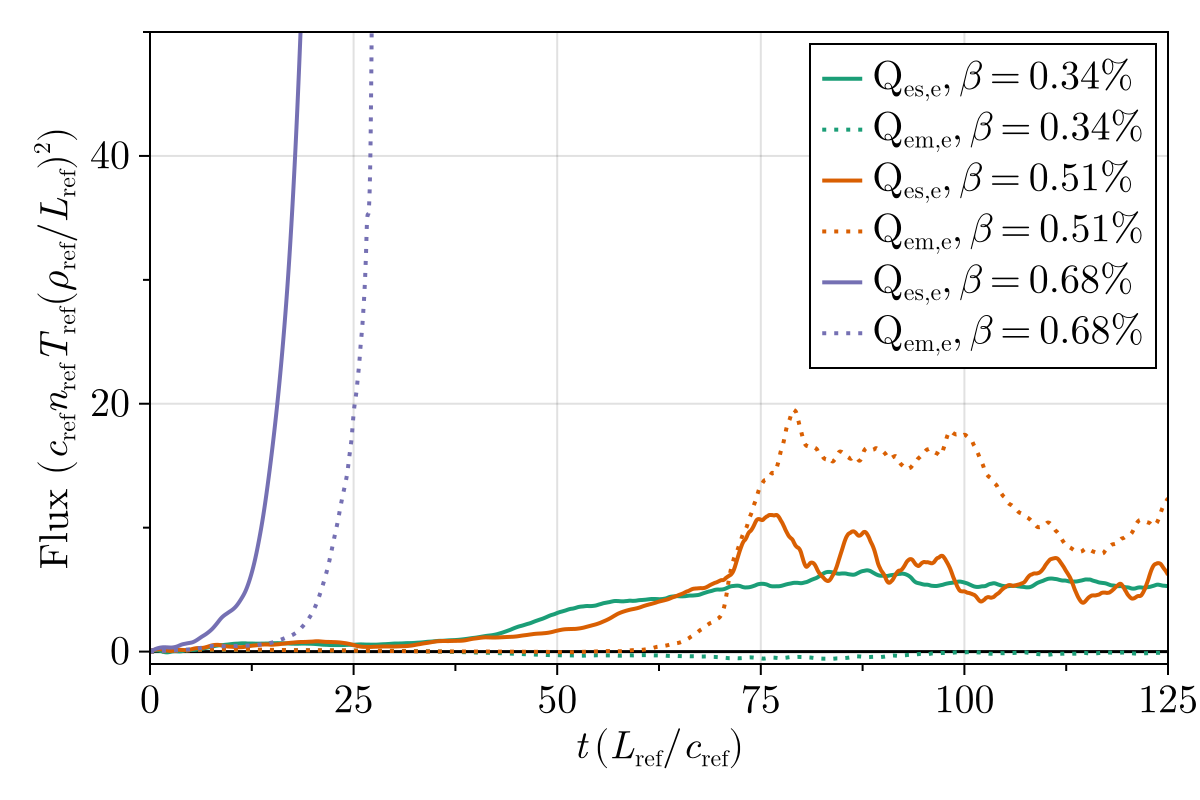}
    \caption{Electron electrostatic and electromagnetic heat fluxes as a function of time from nonlinear \textsc{Gene} simulations at $\beta = \{0.34\%,0.51\%,0.68\%\}$.}
    \label{fig:5}
\end{figure}

\begin{figure}
    \centering
    \includegraphics[width=1.0\linewidth]{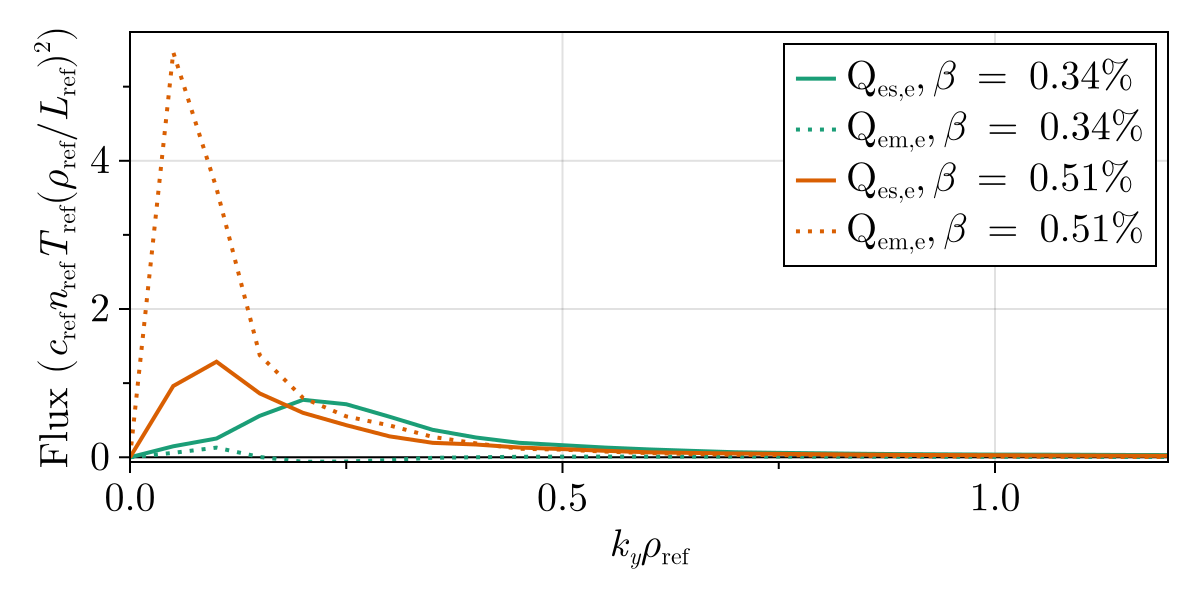}
    \caption{Electron electrostatic and electromagnetic heat fluxes as a function of binormal wavenumber $k_y$ from nonlinear \textsc{Gene} simulations at $\beta = \{0.34\%,0.51\%\}$.}
    \label{fig:6}
\end{figure}

Nonlinear simulations were performed for $\beta = \{0.34\%, 0.51\%, 0.68\%\}$ and present an even more pessimistic picture than the linear analysis.
The time traces of the electron electrostatic and electromagnetic fluxes for these simulations are presented in Figure \ref{fig:5}.
The corresponding decomposition of the fluxes with respect to $k_y$ for the $\beta = \{0.34\%,0.51\%\}$ cases are shown in \ref{fig:6}.
In both simulations, $k_y^\text{min}=0.05$ is used, which has small linear growth rates according to Figure \ref{fig:3}, making it suitable for nonlinear simulations and ensuring that energy is not continuously injected at the system scale.
The linear analysis indicates that $\beta = 0.68\%$ is still in the ITG regime, however, it is clear from Figure \ref{fig:5} that the nonlinear transport grows explosively and does not saturate.
Furthermore, this explosive transport occurs in both the electrostatic and electromagnetic electron transport channels, which is indicative of KBM-driven transport, as the ITG will not drive transport through magnetic field fluctuations.
This presents an even more rigid constraint on $\beta$ than the linear KBM threshold.
For $\beta = 0.51\%$, which is well below the linear KBM threshold, the transport saturates at finite level.
Even so, the electromagnetic transport channel is larger than the electrostatic channel.
Furthermore, as seen in Fig.~\ref{fig:6}, the largest contributions flux are from the lowest $k_y$ wavenumbers, despite these modes having very small linear growth rates in Fig.~\ref{fig:3}.
The strong electromagnetic transport signal indicates that nonlinear excitation can drive large KBM amplitudes, even for modes with almost negligible linear growth rates at parameters below the MHD and KBM $\beta$ thresholds.
This has not been observed previously for quasi-symmetric stellarators but is consistent with findings from W7-X which show that subdominant KBMs are able to drive large electromagnetic transport \citep{mulholland2025}.

%%%%%%%%%%%%%%%%%%%%%%%%%%%%%%%%%%%%%%%%%%%%%%%%%%%%%%%%%%%%%%%%%%%%%%%%%%%%%%%%%%%%%%%%%%%%%%%%%%%%%%%%%%%%%%%%%%%%%%%%%%%%%%%%%%%%%%%%%%%%
\section{Conclusions}\label{sec:4}
This work demonstrates, for the first time, the existence of an explosive transport regime in quasi-axisymmetric stellarators, driven by subdominant, marginally stable kinetic ballooning modes that are strongly excited via nonlinear coupling.
Whereas previous work had shown KMBs to be unimportant up to the global MHD $\beta$ limit for quasi-axisymmetric stellarators \citep{mckinney2021}, we find overwhelmingly deleterious KBM-driven transport well below the linear KBM threshold.
This motivates the need for greater understanding of the underlying KBM dynamics, to better anticipate confinement regime transitions and understand the role of subdominant instabilities that can be excited through nonlinear coupling.

The results presented in this work have significant implications for the design and optimization of next-generation stellarators, including fusion pilot and power plant concepts.
The contrasting picture provided by the macro- and microscopic analyses highlights the importance of multi-scale analysis for design validation.
It also demonstrates the critical need to develop self-consistent reduced transport models and concurrently account for macro- and microscopic instabilities, particularly KBMs, in optimisation procedures.
Finally, our results emphasise the urgent need to address the hard problem of predicting turbulent transport that is driven by subdominant instabilities which, linearly, may be marginally stable.

%%%%%%%%%%%%%%%%%%%%%%%%%%%%%%%%%%%%%%%%%%%%%%%%%%%%%%%%%%%%%%%%%%%%%%%%%%%%%%%%%%%%%%%%%%%%%%%%%%%%%%%%%%%%%%%%%%%%%%%%%%%%%%%%%%%%%%%%%%%%

\begin{acknowledgments}
This research used resources of the National Energy Research Scientific Computing Center (NERSC), a Department of Energy User Facility using NERSC award FES-ERCAP0032166.
\end{acknowledgments}

\bibliography{master}

\end{document}